# Static and Dynamic Oxide-Trapped-Charge-Induced Variability in Nanoscale CMOS Circuits

Gennady Zebrev

*Abstract*—The inter-device mismatch and intra-device temporal instability in the nanoscale CMOS circuits is examined from a unified point of view as a static and dynamic parts of the variability concerned with stochastic oxide charge trapping and de-trapping. This approach has been benchmarked on the recent evidence of the radiation-induced increase of inter-transistor mismatch in 60 nm ICs. A possible reliability limitation in ultrascale circuits concerned with the single or a few charged defect instability is pointed out and estimated.

*Index Terms*—Variability, CMOS, MOSFET, oxide traps, Random Telegraph Noise, 1/f noise, mismatch, total ionizing dose

## I. Introduction

Mature technologies rely upon production of identical copies of some basic components. For example, the digital CMOS technology is based on replication of a large number of preferably identical copies of the MOSFETs. The MOSFET's typical drain current is approximately invariant under the constant aspect ratio size shrinking. This fundamental physical property of 2D devices has provided an opportunity for the aggressive geometrical scaling with an improvement of performance. As the feature size of microelectronic chips is reducing, the variability of electronic components has become a very significant issue for contemporary chips [1, 2, 3]. The major source of static variability is due to the discrete nature of the charged impurities in the devices. For example, the variation in the number and position of the dopant atoms underneath the MOSFET's gates makes each transistor different, introducing device-to-device statistical spread in the device's parameters. Particularly, the random dopant fluctuations in the depletion region of the bulk MOSFETs are the cause of mismatch in their threshold voltage [1]. Modern highly-scaled digital circuits have rather low noise margins and the threshold voltage mismatch could seriously degrade their functionality especially for the ultra-low-power systems, operating in the subthreshold mode [4]. The maximum clock speeds have saturated at < 5GHz partly because of the variance of FETs and circuits at < 40 nm [5].

In contrast to the static inter-transistor variability, the intra-transistor dynamic variability is closely related with the time-dependent drain current fluctuations due to the trapping/de-trapping of carriers in the gate dielectrics. The near-interfacial ("border") traps are responsible for such fluctuations over wide temporal ranges. The impact of the low-frequency and random telegraph noise on the dynamic variability of a single SRAM cell is examined in [6, 7]. The ionizing irradiation is able to significantly enhance the device variability due to the buildup of the random charged defects in the isolation oxides and at the Si-SiO$_2$ interface. Different aspects of such radiation-induced variability have been discussed in [8, 9, 10, 11]. Gerardin et al. investigated experimentally in [12] the interrelations between inter-device static variability and the total dose effects in commercial 65-nm CMOS technology.

The aim of this paper is to consider all these effects from a unified physical point of view.

## II. Model formulation

### A. Screening of external oxide charge

The external oxide charge is screened by the image charges in the channel, substrate and the gate. For the charge $\delta Q_{OX}$, trapped near the silicon (we will denote the total charges and capacitances by the capitalized indices), we have a shift of the surface potential $\varphi_S$ at a fixed gate voltage $V_G$

$$\delta \varphi_S = \frac{\delta Q_{OX}}{C_{OX} + C_D + C_{IT} + C_Q} \qquad (1)$$

where $C_{OX}$ is the gate oxide capacitance, $C_D = dQ_D/d\varphi_S$ is the depletion layer capacitance, $C_{IT}$ is the interface trap capacitance, $C_Q = dQ_C/d\varphi_S$ is the inversion layer ('quantum') capacitance [13]. Due to a strong dependence of $Q_C$ on the surface potential $\varphi_S$ in the subthreshold region, the $C_Q$ does not practically affect the silicon FET gate capacitance $C_G = \left[ C_{OX}^{-1} + \left(C_D + C_{IT} + C_Q\right)^{-1} \right]^{-1}$ (see the inset in Fig. 1), since it is extremely low in the subthreshold operation mode ($C_Q \ll C_D$) and very high in the above threshold strong inversion regime ($C_Q \gg C_D$). Under such circumstances, the quantum capacitance could be well estimated in practice in a non-degenerate approximation [14]

$$C_Q \cong \frac{Q_C}{2\varphi_T}\left(1 + \frac{Q_D}{Q_C + Q_D}\right), \qquad (2)$$

G. I. Zebrev is with the Department of Micro- and Nanoelectronics of National Research Nuclear University MEPHI, Moscow, Russia (e-mail: gizebrev@mephi.ru).

where $\varphi_T = k_B T/q$ is the thermal potential, $Q_C$, $Q_D$ are the total charge in the inversion and depletion layers respectively. Provided a weak dependence of the carrier's mobility on $V_G$, the inverse logarithmic slope can be calculated as a function of the relevant capacitances and the channel charge in the following way

$$S \equiv I_D \frac{dV_G}{dI_D} \cong Q_C \frac{dV_G}{dQ_C} = \left(\frac{Q_C}{dQ_C/d\varphi_S}\right)\frac{dV_G}{d\varphi_S} = \\ = \varphi_D\left(1 + \frac{C_D + C_{IT} + C_Q}{C_{OX}}\right) = \varphi_D\left(1 + \frac{C_D + C_{IT}}{C_{OX}}\right) + \frac{Q_C}{C_{OX}}, \quad (3)$$

where the diffusion potential $\varphi_D$ is defined and could be estimated as follows

$$\varphi_D \equiv \frac{Q_C}{dQ_C/d\varphi_S} = \frac{Q_C}{C_Q} \cong 2\varphi_T\left(1 + \frac{Q_D}{Q_C + Q_D}\right)^{-1}. \quad (4)$$

As can be seen in (3) an absolute value of $\varphi_D$ is important only in the subthreshold mode, where $\varphi_D \cong \varphi_T$.

It is instructive to consider $S$ in two limiting cases. First, in the subthreshold region ($Q_C \ll Q_D$) it corresponds to the well-known subthreshold slope measured in Volts per decade of gate voltage

$$SS = S\ln 10 = \varphi_T \ln 10\left[1 + (C_D + C_{IT})/C_{OX}\right] \equiv m\varphi_T \ln 10, \quad (5)$$

where $m$ is often referred to as an ideality factor. Second, in the strong inversion region, we have $S \cong Q_C/C_{OX} \propto V_G - V_T$. The inverse logarithmic slope $S$ is closely related with the transconductance $g_m \equiv dI_D/dV_G$

$$S = I_D(dV_G/dI_D) = I_D/g_m. \quad (6)$$

Fig. 1 shows the physical meaning of $S$ and illustrates its interrelation with $g_m$.

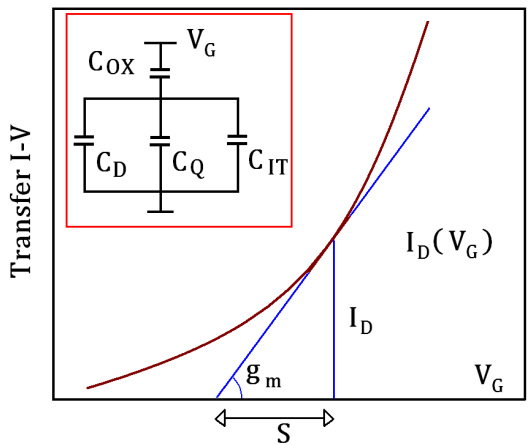

Fig. 1. Field-effect transistor transfer characteristics and graphical representation of the transconductance $g_m$ and the inverse logarithmic slope $S$.

Thus defined logarithmic slope describes in a unified way both the strong inversion and the weak inversion regions. In contrast to the extensive transconductance (i.e., dependent on $W/L$), the inverse logarithmic slope $S$ is a thermodynamically intensive variable, i.e., independent of the channel size and shape.

The quantum capacitance $C_Q$ determines the channel charge fluctuation $\delta Q_C = C_Q \delta \varphi_S$ in all operation modes. Using (1)-(4), one could obtain a general relation

$$\frac{\delta Q_C}{Q_C} = \frac{C_Q \delta \varphi_S}{Q_C} = \frac{\delta Q_{OX}}{C_{OX} S} \quad (7)$$

In the strong inversion mode, where $C_Q \gg C_{OX}, C_D, C_{IT}$, we have very small $\delta \varphi_S$, $S \cong Q_C/C_{OX}$ and the "perfect screening" by the channel $\delta Q_C/\delta Q_{OX} \leq 1$. This is due to the fact that an external oxide charge is screened in the strong inversion on the low spatial scales of order of the inter-electron distance in the channel. For the sake of brevity, we ignore here the effects of the accidental electrostatic coupling with the source/drain contacts which could reduce the image charge $\delta Q_C$, especially in the short channels.

Using (7), one can obtain the relation

$$\frac{\mathrm{var}\, Q_C}{Q_C^2} = \frac{\mathrm{var}\, Q_{OX}}{C_{OX}^2 S^2} = \frac{\mathrm{var}\, V_T}{S^2}, \quad (8)$$

where $\mathrm{var}\, Q_{OX}$ and $\mathrm{var}\, Q_C$ are the variances of the oxide and the channel charge fluctuations, and $\mathrm{var}\, V_T = \mathrm{var}\, Q_{OX}/C_{OX}^2$ is the variance of the threshold voltage $V_T$. This result can be used for analysis of the static and the dynamic kinds of the oxide-trapped-induce variability in nanoscale MOSFETs.

### III. MISMATCH AS STATIC VARIABILITY

#### A. Static variability of the threshold voltage

Equation (8) allows describing the sample-to-sample static drain current fluctuations for all MOSFET's operational modes. Actually, taking into account (8), one gets

$$\frac{\mathrm{var}\, I_D}{I_D^2} \cong \frac{\mathrm{var}\, Q_C}{Q_C^2} = \frac{\mathrm{var}\, V_T}{S^2} = \frac{g_m^2 \mathrm{var}\, V_T}{I_D^2}, \quad (9)$$

where the threshold voltage variance is assumed to be distributed by a set of transistors. For the above threshold mode ($V_G > V_T$) we have $S \cong Q_C/C_{OX} \sim V_G - V_T$ and typically low levels of the drain current inter-device mismatch $\mathrm{var}\, I_D/I_D^2 \sim \mathrm{var}\, V_T/(V_G - V_T)^2 \ll 1$, while the subthreshold mode ($V_G < V_T$) corresponds to $S \cong \varphi_T m$, that could provide a wide drain current spread $\delta I_D/I_D > 1$.

The variations of the threshold voltage $u_T$ around their mean values $V_T$ are normally described in MOSFETs by the Gauss distribution

$$P_{V_T}(u_T) = \frac{1}{\left[2\pi\, \mathrm{var}\, V_T\right]^{1/2}} \exp\left(-\frac{(u_T - V_T)^2}{2\, \mathrm{var}\, V_T}\right). \quad (10)$$

The threshold voltage variance $\mathrm{var}\, V_T$ can be calculated summing up presumably independent terms, corresponding to the dopant atoms in the silicon substrate and to the radiation-induced charged traps in the oxide

$$\operatorname{var} V_T = \left(\frac{\partial V_T}{\partial Q_D}\right)^2 \langle \delta Q_D^2 \rangle + \left(\frac{\partial V_T}{\partial Q_{OX}}\right)^2 \langle \delta Q_{OX}^2 \rangle =$$
$$= \frac{\operatorname{var} Q_D + \operatorname{var} Q_{OX}}{C_{OX}^2} \equiv \operatorname{var} V_T^{RDF} + \operatorname{var} V_T^{OX}, \quad (11)$$

where $\operatorname{var} Q_D$ and $\operatorname{var} Q_{OX}$ are the variances of the random dopant atom (RDF) and the charged oxide trap numbers.

Since the variance equals the average for the Poisson distribution, the dopant charge fluctuation can be written as $\langle \delta Q_D^2 \rangle = q Q_D$ and then we have an expression for the RDF part of the threshold voltage variance

$$\operatorname{var} V_T^{RDF} \cong \frac{q Q_D}{C_{OX}^2} = \frac{q^2}{(\varepsilon_0 \varepsilon_{ox})^2} \left(\frac{4\varphi_F \varepsilon_0 \varepsilon_S N_A}{q}\right)^{1/2} \frac{t_{ox}^2}{WL}, \quad (12)$$

where $N_A$ is a doping level of the p-Si substrate, $\varphi_F = \varphi_T \ln N_A / n_i$ is the bulk Fermi potential, $n_i$ is the Si intrinsic concentration, $\varepsilon_0 \varepsilon_{ox}$ and $t_{ox}$ are the gate insulator's permittivity and thickness. The RDFs are significantly suppressed in modern 3D FET configurations and typically unaffected by impacts of ionizing irradiation, hot electrons or other non-equilibrium external influence.

The charge trapping in the oxide is a stochastic factor concerned with an impact of ionizing radiation, single event radiation effects in space, or in other hazardous environments. The threshold voltage shift under irradiation is determined by the net oxide charge

$$\Delta V_T = -\left(Q_{OX}^+ - Q_{OX}^-\right)/C_{OX}. \quad (13)$$

where $Q_{OX}^+$ ($Q_{OX}^-$) is an effective amount of the positive (negative) charge trapped near the Si-SiO$_2$ interface. The positive radiation-induced oxide-trapped charge is often (especially under the low dose rate irradiation) strongly compensated due to the tunnel relaxation and/or interface trap buildup in the n-MOSFETs [15]. The very thin gate oxides also make $\Delta V_T$ negligible even at rather high doses. At the same time, the variance of the oxide charge number is described not by a net charge but by a sum of the charged defects with different signs $\operatorname{var} Q_{OX} = q\left(Q_{OX}^+ + Q_{OX}^-\right) \equiv q^2 N_{OX}$. Then we have

$$\operatorname{var} V_T^{OX} = \operatorname{var} Q_{OX} / C_{OX}^2 = q^2 N_{OX} / C_{OX}^2. \quad (14)$$

Thus, the ionizing radiation may have a little effect on the average I-V characteristics of modern MOSFETs, greatly increasing at the same time the spread of their parameters.

### B. Lognormal current distribution in subthreshold modes

The drain current in the subthreshold region ($V_G < V_T$) of MOSFET with a random threshold voltage $u_T$ is well described by a simple exponential approximation

$$i_D(u_T) \cong I_T \exp\left(\frac{V_G - u_T}{S}\right), \quad (15)$$

where the subthreshold slope is expressed via a constant ideality factor $S \cong m \varphi_T$. The Gaussian (normal) distribution of the threshold voltages (10) in this mode is transformed into a lognormal distribution of the subthreshold drain currents

$$P_I(i_D) = P_{V_T}\left[V_T(i_D)\right]\left|\frac{dV_T}{dI_D}\right| =$$
$$= \frac{m \varphi_T}{\sqrt{2\pi \operatorname{var} V_T} \, i_D} \exp\left(-\frac{\operatorname{var} V_T}{2 m^2 \varphi_T^2} \ln^2\left(i_D/\bar{I}_D\right)\right), \quad (16)$$

where $\bar{I}_D$ is a medial parameter of the lognormal distribution, corresponding to a mean threshold voltage $\bar{I}_D = I_T \exp\left[(V_G - V_T)/m \varphi_T\right]$. The subthreshold drain current averaged over an ensemble of $N$ transistors with the scattered threshold voltages [16] can be calculated as an averaged over a lognormal distribution

$$\langle I_D \rangle = N^{-1} \sum_{i=1}^{N} I_D^{(i)} = \int i_D P_I(i_D) di_D = \bar{I}_D \exp\left(\frac{\operatorname{var} V_T}{2 m^2 \varphi_T^2}\right). \quad (17)$$

Fig. 2 shows the shapes of current distribution calculated at a fixed current with different values of $\operatorname{var} V_T$.

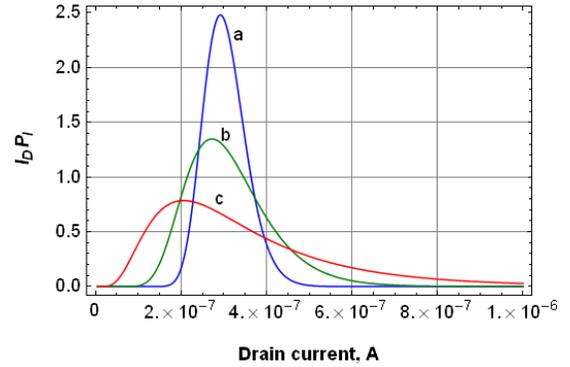

Fig. 2. Drain current distributions $I_D P_I(I_D)$ calculated at a fixed $\bar{I}_D = 3 \times 10^{-7}$ A ($W/L$=120 nm/60 nm, $N_A = 10^{17}$ cm$^{-3}$) at (a) $N_{ox} = 0$; (b) $N_{ox} = 10^{12}$ cm$^{-3}$ ; (a) $N_{ox} = 2 \times 10^{13}$ cm$^{-3}$. A feasible shift of the threshold voltage is set to be zero.

Strictly speaking, the applicability of the Gaussian and lognormal distributions is valid only for the large area devices. To take into account a discrete nature of the trapped charge, one has to average the drain current over Poisson's distributions with the expected numbers for the trapped charge of both signs $Q_{OX}^{+/-} = q \langle n \rangle_{+/-}$

$$\langle I_D \rangle =$$
$$= I_T e^{\frac{V_G - V_{T0}}{m \varphi_T}} \sum_{n=0}^{\infty} \sum_{m=0}^{\infty} \frac{\langle n \rangle_+^n}{n!} \frac{\langle n \rangle_-^m}{m!} e^{-\langle n \rangle_+ - \langle n \rangle_-} \exp\left[\frac{q(n-m)}{C_{OX} m \varphi_T}\right] = \quad (18)$$
$$= I_T e^{\frac{V_G - V_{T0}}{m \varphi_T}} \exp\left[\frac{Q_{OX}^+}{q}\left(e^{\frac{q}{C_{OX} m \varphi_T}} - 1\right) + \frac{Q_{OX}^-}{q}\left(e^{-\frac{q}{C_{OX} m \varphi_T}} - 1\right)\right].$$

Expanding an exponent in powers of $q/C_{OX} m \varphi_T$ (typically $\ll 1$) to a square term, one gets the result

$$\langle I_D \rangle \cong I_T e^{\frac{V_G - V_{T0}}{m \varphi_T}} \exp\left[\frac{\left(Q_{OX}^+ - Q_{OX}^-\right)/C_{OX}}{m \varphi_T} + \frac{q\left(Q_{OX}^+ + Q_{OX}^-\right)/C_{OX}^2}{2 m^2 \varphi_T^2}\right], \quad (19)$$

that, in view of $V_T = V_{T0} - \left(Q_{OX}^+ - Q_{OX}^-\right)/C_{OX}$, is essentially the same as (17). The average drain current exceeds the medial value due to the current distribution in the subthreshold region is skewed to the right at a sufficiently large $\text{var} V_T$. This means that a relatively small portion of transistors with negatively shifted $V_T$ provides a significant contribution to average current because of strong dependence of drain current on $V_G$ in the subthreshold region.

*C. Numerical simulation of inter-device fluctuations of static drain currents*

Characterization of the drain current variability can be generally addressed as the ratio of the standard deviation to the mean current $\left(\langle I_D^2\rangle - \langle I_D\rangle^2\right)^{1/2}/\langle I_D\rangle$, calculated via a straightforward averaging of the I-V characteristics. We have numerically simulated the drain current variance at any operation mode using a general formula

$$\text{var} I_D(V_G) = \int I_D^2(V_G - u_T) P_{V_T}(u_T) du_T - \left[\int I_D(V_G - u_T) P_{V_T}(u_T) du_T\right]^2, \quad (20)$$

where the threshold voltage variance can be specified for different channel sizes, doping levels, and oxide-trapped charges. Such approach requires an analytical dependence of the drain current on gate and drain biases for all operation modes. We use in this case the compact MOSFET model, described in [17]. Figure 3 shows the standard deviation to the mean value drain current ratio simulated as functions of the gate voltage. The amplitudes of the current standard deviations are significantly larger in the subthreshold operation modes due to lesser values of the inverse logarithmic slope $S$ in (8).

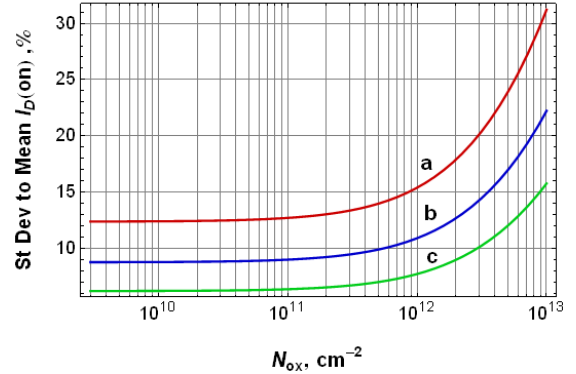

Fig. 4. The standard deviation to mean values of on-current ($V_G = V_{DD}$=1.2 V) simulated as functions of trap concentration $N_{ox}$ at different sizes (a) $W/L$=60 nm/60 nm (b) $W/L$=120 nm/60 nm (c) $W/L$=240 nm/60 nm for the same nFET as in Fig. 3.

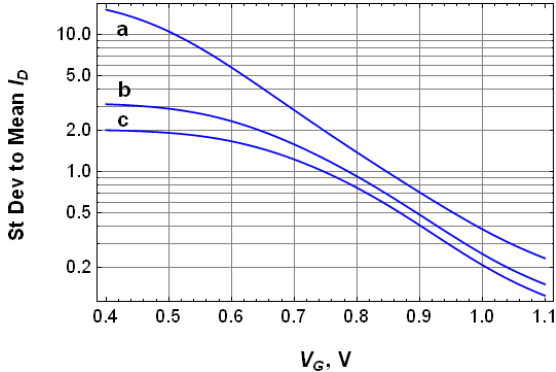

Fig. 3. The drain current standard deviation to mean values simulated as functions of gate voltage at different oxide trap concentrations $N_{ox}$: (a) $10^{13}$ cm$^{-3}$ (b) $3\times 10^{12}$ cm$^{-3}$ (c) $10^{11}$ cm$^{-3}$ for a 60 nm nFET ($t_{ox}$ = 3 nm, $W/L$=120 nm/60 nm, $V_T$ = 0.7 V, $N_A$ = $3\times 10^{17}$ cm$^{-3}$).

The normalized standard deviation of the on-current (at $V_G = V_{DD}$) has to be an increasing function of $N_{ox}$. This is illustrated by the simulation results in Fig. 4.

For the low oxide charge density, the relative current fluctuations are constant due to the dominance of the RDF in (11). The height of the plateau in Fig. 4 is determined by $\text{var} V_T^{RDF}$ in (12). A noticeable increase in the drain current fluctuations at large $N_{ox}$ is caused by fluctuations of the oxide charge $\text{var} V_T^{ox}$.

Thus, ionization may cause additional inter-device threshold voltage mismatch. The trapped charge compensation could strongly suppress the net threshold voltage shift making it sublinear or negligible for contemporary nanoscale circuits with the thin gate oxides. At the same time, the total density of the charged oxide traps $N_{ox} \equiv N_{ox}^+ + N_{ox}^-$ could be considered as a function approximately proportional to the total ionizing dose. Particularly, the dose dependence of the relative drain current fluctuations, measured for the 120/60 nm nMOSFETs (see Fig. 6 in Ref. [12]), is almost identical with the curve (b) in Fig. 4, intentionally simulated with the authentic transistor's parameters. Evidently, the severity of this problem would increase with a shrinking of transistor sizes.

IV. DYNAMIC VARIABILITY

*A. Dynamic variability*

Dynamic variability can be considered as a form of dynamic mismatch. Actually, the oxide-trapped charged is not fixed due to carrier exchange with the silicon substrate. Following the changes in the Fermi energy position at the Si-SiO$_2$ interface, the near-interfacial charged defects could change their occupation number (trapping/de-trapping), contributing both to the interface trap capacitance and the threshold voltage instability [18, 19, 20].

The trap response times have a very wide range and delayed kinetics. The auto-correlator of the dynamical threshold voltage can be derived for uniformly distributed traps in approximation of a stationary temporal process [21, 22]

$$K_{V_T}(\Delta t) = \langle \delta V_T(0) \delta V_T(\Delta t) \rangle = \frac{\text{var} Q_r}{C_{OX}^2} \frac{\lambda}{\ell} \left(E_1\left(\frac{\Delta t}{\tau_{max}}\right) - E_1\left(\frac{\Delta t}{\tau_{min}}\right)\right), \quad (21)$$

where $\text{var} Q_r$ is the variance of total number of the traps $Q_r$ recharged per gate voltage sweep averaged over all temporal scales, $E_1(y)$ is the integral exponent function ($E_1(y) \cong -0.577 - \ln y$ at $y \ll 1$), the maximum and minimum times of the tunneling recharging are related as $\tau_{max} = \tau_{min} \exp(\ell/\lambda)$, $\ell$ is a thickness of the trap location near the Si-SiO$_2$ interface (typically, a few nanometers), $\lambda$ is the tunnel attenuation length ($\leq 0.1$ nm).

The quasi-static interface trap capacitance $C_{IT}$ and the trap energy density $D_{IT}$ are defined as follows

$$C_{IT} = -q dQ_{OX}/d\varepsilon_F = q^2 D_{IT}. \qquad (22)$$

The total variance $\mathrm{var}\, Q_r$ at a single voltage sweep can be estimated as

$$\mathrm{var}\, Q_r = q \overline{C}_{IT} \Delta \varphi_S, \qquad (23)$$

where $\Delta\varphi_S \cong \varphi_T \ln N_A/n_i$ is a typical interval of the surface potential (or Fermi energy) change under gate sweep, $\overline{C}_{IT}$ is the interface trap capacitance averaged over this interval.

Such a spread in $Q_r$ could lead to a variety of measured threshold voltages. In contrast to the static mismatch, the dynamical variability of threshold voltages essentially depends on the sweep time $t_S$. All the high-frequency fluctuations are self-averaging over time scales lesser than $t_S$. Given $t_S \ll \tau_{\max}$, equation (21) takes an asymptotic form

$$K_{V_T}(t_S) \cong \frac{q\overline{C}_{IT}\Delta\varphi_S}{C_{OX}^2}\left(1 - \frac{\lambda}{\ell}\ln\frac{t_S}{t_1}\right), \qquad (24)$$

where $t_1$ is the reference time scale. We found in [21] that $t_1$, $\overline{C}_{IT}$ and $\ell$ are not independent parameters, and they can be consistently recalculated with a renormalization procedure depending on the experimental conditions.

Actually, as it was experimentally found in [19, 23], the dynamic standard deviation of the threshold voltage approximately logarithmically decreases with increase in the sweep time $t_S$. Such single device dynamic variability could be added to the static mismatch contribution and could amount up to ≈30% of static variability sources. The lack of dynamic stability in the nanoscale memory circuits can lead to read/write failures or power supply limitations.

*B. Flicker noise*

The 1/$f$, or flicker noise and the random telegraph noise belong to a class of intra-device variability defined as time-dependent fluctuations of the drain current around its fixed average value. In fact, the flicker noise is due to the same processes as dynamical variability. Indeed, in view of (9) and (21), we derived with the McWhorter model [24] the current auto-correlation noise function

$$\frac{S_{I_D}(\Delta t)}{I_D^2} = \frac{qC_{IT}(\varepsilon_F)\varphi_T}{C_{OX}^2 S^2}\frac{\lambda}{\ell}\left(E_1\left(\frac{\Delta t}{\tau_{\max}}\right) - E_1\left(\frac{\Delta t}{\tau_{\min}}\right)\right), \qquad (25)$$

where the dispersion of trapped oxide charge is controlled by the interface (border) trap energy density at a fixed Fermi level $C_{IT}(\varepsilon_F)$ [25]. According to the Wiener-Khinchine theorem, the power spectral density is defined by a cosine-transform of the auto-correlation function

$$\frac{S_{I_D}(\omega)}{I_D^2} = \frac{4\int_0^\infty S_{I_D}(t)\cos(\omega t)dt}{I_D^2} =$$
$$= 4\frac{qC_{IT}(\varepsilon_F)\varphi_T}{C_{OX}^2 S^2}\frac{\lambda}{\ell}\frac{\tan^{-1}(\omega\tau_{\max}) - \tan^{-1}(\omega\tau_{\min})}{\omega} \cong$$
$$\cong \frac{qC_{IT}(\varepsilon_F)\varphi_T}{C_{OX}^2 S^2 \ln(\tau_{\max}/\tau_{\min})}\frac{1}{f}. \qquad (26)$$

It is worth to note that the thermal relaxation time constants can have a huge scatter in magnitude also due to a spread in the activation energy $\varepsilon_{\max} > \varepsilon > \varepsilon_{\min}$, $\tau_0 e^{\varepsilon_{\max}/k_B T} > \tau > \tau_0 e^{\varepsilon_{\min}/k_B T}$. The mathematical structure of the response function for the tunnel ("horizontal") and the thermal ("vertical") relaxation for the traps, uniformly distributed in position and energy, is equivalent each other up to a substitution $k_B T/(\varepsilon_{\max}-\varepsilon_{\min}) \leftrightarrow \lambda/\ell \leftrightarrow \ln(\tau_{\max}/\tau_{\min})$ [21]. It is well known that the 1/$f$ noise is a superposition of the Random Telegraph Noise (RTN) originating from the trapping/de-trapping of the separate defects [26]. Besides, the Negative Bias Temperature Instability (NBTI) is found to be caused by the same reasons [27, 28, 29].

*C. Ultimately scaled circuits*

The stochastic oxide-trapped charge could have a huge impact on the reliability of the ultimately scaled circuit. The drive current in transistors of contemporary technologies can be estimated as follows

$$I_D = \frac{Q_C}{L} v_{\max} \qquad (27)$$

where $v_{\max}$ is the maximum carrier speed (~ $10^7$ cm). Then the relative current fluctuation is given by

$$\frac{\delta I_D}{I_D} = \frac{\delta Q_{OX}}{C_{OX} S} \cong \begin{cases} \dfrac{\delta Q_{OX}}{C_{OX}(V_G - V_T)}, & V_G > V_T, \\ \dfrac{\delta Q_{OX}}{m\varphi_T}, & V_G < V_T. \end{cases} \qquad (28)$$

For transistors with 10nm×10nm sizes, the total channel charge comprises of several electrons even at maximum gate voltages. The trapping/de-trapping process even in a single defect ($\delta Q_{ox} \geq q$) could lead in this case to a noticeable variation in drive current of such transistors. These fluctuations become critical on the condition $\delta Q_{ox} \sim q \sim C_{OX} S$. The critical channel area for strong inversion can be estimated via the equivalent oxide thickness (EOT) as follows

$$A_{crit} \sim \frac{q}{C_{ox}(V_{DD}-V_T)} = \frac{q\, EOT}{\varepsilon_{ox}(V_{DD}-V_T)}. \qquad (29)$$

Taking $V_{DD} - V_T = 0.5$ V and EOT = 1 nm, one gets $A_{crit} \cong 9$ nm$^2$. In the subthreshold region the impact of an individual charged oxide trap should be noticeable even for modern 10 nm technology

$$A_{crit} \sim \frac{q\, EOT}{\varepsilon_{ox}\varphi_T} \sim 180 \text{ nm}^2. \qquad (30)$$

We claim in this way that $A_{crit} \sim 10$ nm$^2$ is the ultimately minimal channel area, limited by unavoidable variability due to stochastic charge instability of a single defect in surrounding insulators.

## V. Summary

The static and dynamic variability, flicker noise and RTN in nanoscale CMOS has been treated in this work from a single standpoint. A distinctive general feature of all these effects is the Pelgrom's dependence $\propto 1/\sqrt{WL}$ for the standard deviation of the amplitude [30]. This dependence is a direct consequence of the Poisson statistics for the charged defects distributed without correlations on the transistor's channel area $A = WL$. For the channel area less than approximately 10 nm$^2$ the static and dynamic variability effects may become unacceptably large due to the fluctuations may be comparable with the average values. The advanced technologies (e.g., FinFETs, or, FD SOI FETs) allow excluding or minimizing the RDF effect, but no technology can prevent the emergence of several unstable defects in even a very thin surrounding insulator. This issue represents a fundamental limit on the reliability of the ultra-scale devices and circuits.

REFERENCES

[1] K. Bernstein, D. J. Frank, *et al.*, "High-Performance CMOS Variability in the 65-nm Regime and Beyond," *IBM J. Res. & Dev.*, Vol. 50, No. 4/5, pp. 433–449, July/September 2006, doi: 10.1147/rd.504.0433

[2] C. Mezzomo, A. Bajolet, A. Cathignol, R. Di Frenza, and G. Ghibaudo, "Characterization and modeling of transistor variability in advanced CMOS technologies," *IEEE Trans. Electron Devices*, vol. 58, no. 8, pp. 2235–2248, Aug. 2011, doi: 10.1109/TED.2011.2141140

[3] B. Nikolic, Ji-Hoon Park, Jaehwa Kwak, B. Giraud, Zheng Guo, Liang-Teck Pang, Seng Oon Toh, R. Jevtic, Kun Qian and C. Spanos, "Technology Variability From a Design Perspective," *IEEE Trans. Circuits and Systems I:Regular Papers,.*, Vol. 58, pp. 1996-2009, 2011, doi: 10.1109/TCSI.2011.2165389

[4] Alice Wang, Benton H. Calhoun, Anantha P. Chandrakasan, *Sub-threshold Design for Ultra Low-Power Systems*, Springer, USA, 2006.

[5] *Chips 2020*, Vol. 2, ed. by B. Hoefflinger, Springer, 2016, doi: 10.1007/978-3-319-22093-2

[6] C. G. Theodorou, M. Fadlallah, Xavier Garros, C. Dimitriadis, G. Ghibaudo, "Noise-induced dynamic variability in nano-scale CMOS SRAM cells," ESSDERC, Lausanne, Switzerland, 2016, doi: 10.1109/ESSDERC.2016.7599634

[7] E.G. Ioannidis, S. Haendler, J.-P. Manceau, C. A. Dimitriadis and G. Ghibaudo, "Impact of dynamic variability on the operation of CMOS inverter," *Electronics Letters*, Vol.49, No.19, pp. 1214-1216, 2013, doi: 10.1049/el.2013.1343

[8] Y. Li, N. Rezzak, E. X. Zhang, R. D. Schrimpf, D. M. Fleetwood, J. Wang, D. Wang, Y. Wu, and S. Cai, "Including the effects of process related variability on radiation response in advanced foundry process design kits," *IEEE Trans. Nucl. Sci.*, Vol. 57, No. 6, pp. 3570–3574, Dec. 2010, doi: 10.1109/TNS.2010.2086478.

[9] M. S. Gorbunov, I. A. Danilov, G. I. Zebrev, P.N. Osipenko, "Verilog-A modeling of radiation-induced mismatch enhancement," *IEEE Trans. Nucl. Sci.,* Vol. 58, No. 3, pp. 785-792, June 2011, doi: 10.1109/TNS.2010.2104162

[10] G. I. Zebrev, M. S. Gorbunov, "Radiation Induced Leakage Due to Stochastic Charge Trapping in Isolation Layers of Nanoscale MOSFETs," *Proceedings SPIE*, Vol. 7025, P. 702517-702517-8, 2008, doi: 10.1117/12.802480

[11] E. Chatzikyriakou, W. Redman-White, C.H. De Groot, "Total Ionizing Dose, Random Dopant Fluctuations, and its combined effect in the 45 nm PDSOI node," *Microelectron. Reliability*, Vol. 68, pp.21-29, Jan. 2017, doi: 10.1016/j.microrel.2016.11.007.

[12] S. Gerardin, M. Bagatin, D. Cornale, L. Ding, S. Mattiazzo, A. Paccagnella, F. Faccio, and S. Michelis, "Enhancement of Transistor-to-Transistor Variability Due to Total Dose Effects in 65-nm MOSFETs," *IEEE Trans. Nucl. Sci*., Vol. 62, No. 6, pp. 2398–2403, Dec. 2015, doi: 10.1109/TNS.2015.2498539.

[13] S. Luryi, "Quantum Capacitance Devices," *Appl. Phys. Lett.*, Vol. 52, No.6, pp. 501 503, Feb. 1988, doi: 10.1063/1.99649

[14] G. I. Zebrev, R. G. Useinov, "Simple Model of Current-Voltage Characteristics of a Metal-Insulator-Semiconductor Transistor," *Sov. Phys. Semiconductors*, Vol. 24, No. 5, pp. 491–493, May 1990, WOS: A1990EK96700001.

[15] T.-P. Ma, P. V. Dressendorfer (Eds), *Radiation Effects in MOS Devices and Integrated Circuits*, John Wiley & Sons, 1988.

[16] D. J. Frank, R. H. Dennard, E. Nowak, P. M. Solomon, Y. Taur, and H.-S. P. Wong, ''Device Scaling Limits of Si MOSFETs and Their Application Dependencies,'' *Proc. IEEE,* Vol. 89, No. 3, pp. 259–288, 2001, doi: 10.1109/5.915374

[17] G. I. Zebrev, V. V. Orlov, A. S. Bakerenkov, V. A. Felitsyn, "Compact Modeling of MOSFET's I-V Characteristics and Simulation of Dose-Dependent Drain Current," *IEEE Trans. Nucl. Sci.,* Vol. 64, No. 8, pp. 2212-2218, Aug. 2017, doi: 10.1109/TNS.2017.2712284.

[18] E. G. Ioannidis, S. Haendler, C. G. Theodorou, N. Planes, C. A. Dimitriadis, G. Ghibaudo, "Statistical Analysis of Dynamic Variability in 28nm FD-SOI MOSFETs," doi:10.1109/ESSDERC.2014.6948798.

[19] K. Takeuchi, T. Nagumo, K. Takeda, et al., "Direct observation of RTN-induced SRAM failure by accelerated testing and its application to product reliability assessment," *IEEE Proc. Symp. VLSI Technology*, Honolulu, June 2010, pp. 189−190, doi: 10.1109/VLSIT.2010.5556222

[20] V.V. Emeliyanov, G. I. Zebrev, et al., "Reversible positive charge annealing in MOS transistor during variety of electrical and thermal stresses," *IEEE Trans. Nucl. Sci.*, Vol. 43, No. 3, pp. 805–809, Aug. 1996, doi: 10.1109/23.510716

[21] G. I. Zebrev, M. G. Drosdetsky, "Temporal and Dose Kinetics of Tunnel Relaxation of Non-Equilibrium Near-Interfacial Charged Defects in Insulators," *IEEE Trans. Nucl. Sci.*, Vol. 63, No. 6, pp. 2895–2902, Dec. 2016, doi: 10.1109/TNS.2016.2619060

[22] G. I. Zebrev, A. A. Tselykovskiy, E. V. Melnik, "Interface Traps In Graphene Field Effect Devices: Extraction Methods and Influence on Characteristics," in *Graphene Science Handbook: Size-Dependent Properties*, CRC Press, Taylor and Francis Group, LLC, 2016, pp. 145-158, ISBN 13:978-1-4665-9136-3

[23] E. G. Ioannidis, S. Haendler, J.-P. Manceau, C. A. Dimitriadis, G. Ghibaudo, "Impact of dynamic variability on the operation of CMOS inverter," *Electronics Letters*, Vol. 49, No. 19, pp. 1214-1216, 2013, doi: 10.1049/el.2013.1343

[24] A. L. McWhorter, Semiconductor Surface Physics, Ed. by R. H. Kingston, p. 207, University of Pennsylvania Press (1957).

[25] D. M. Fleetwood, "1/f Noise and Defects in Microelectronic Materials and Devices," *IEEE Trans. Nucl. Sci*., Vol. 62, No. 4, pp. 1462–1486, Aug. 2015, doi: 10.1109/TNS.2015.2405852

[26] M. J. Kirton and M. J. Uren, "Noise in solid-state microstructures: A new perspective on individual defects, interface states, and low-frequency noise," *Adv. Phys.*, Vol. 38, No. 4, Nov. 1989, pp. 367-468, doi: 10.1080/00018738900101122

[27] B. Kaczer, T. Grasser, Ph. J. Roussel, J. Franco, R. Degraeve, L.-A. Ragnarsson, E. Simoen, G. Groeseneken, H. Reisinger, "Origin of NBTI Variability in Deeply Scaled pFETs," *2010 IEEE International Reliability Physics Symposium*, 2010, pp. 26–32, doi: 10.1109/IRPS.2010.5488856

[28] P. Weckx, B. Kaczer, M. Toledano-Luque, T. Grasser, Ph. J. Roussel, H. Kukner, P. Raghavan, F. Catthoor, G. Groeseneken "Defect-based Methodology for Workload-dependent Circuit Lifetime Projections – Application to SRAM," *2013 IEEE International Reliability Physics Symposium (IRPS)*, pp. 3A.4.1-3A.4.7, doi: 10.1109/IRPS.2013.6531974

[29] T. Grasser "Stochastic charge trapping in oxides: From random telegraph noise to bias temperature instabilities," *Microelectron. Reliability*, Vol. 52, pp. 39-70, 2012, doi: 10.1016/j.microrel.2011.09.002

[30] M. J. Pelgrom, A. C. Duinmaijer, A. P. Welbers, "Matching properties of MOS transistors," *IEEE Journal of Solid-State Circuits*, Vol. 24. No. 5, pp. 1433–1439, Oct. 1989, doi: 10.1109/JSSC.1989.572629